\documentclass{ifacconf}
\usepackage[usenames,dvipsnames]{xcolor}
\usepackage{graphicx}
\usepackage{natbib}
\usepackage{amssymb}
\usepackage{amsmath}
\usepackage{mathtools}
\usepackage{eurosym}
\usepackage{derivative}
\usepackage{threeparttable}
\usepackage{booktabs}
\usepackage{bm}
\usepackage[FIGTOPCAP]{subfigure}
\usepackage[short]{optidef}
\usepackage[version=3]{mhchem}
\usepackage{float}
\usepackage{dirtytalk}
\newcommand{\tran}{{\mkern-1.5mu\mathsf{T}}}

\begin{document}
\begin{frontmatter}

\title{Linking intra- and extra-cellular metabolic domains via neural-network surrogates for dynamic metabolic control}

\thanks[footnoteinfo]{This work was supported by the U.S. Department of Energy, Office of Science, Office of Biological and Environmental Research (Award Number DE-SC0022155).}

\author[First]{Sebastián Espinel-Ríos} 
\author[First,Second,Third,Fourth]{José L. Avalos}

\address[First]{Department of Chemical and Biological Engineering, Princeton University, United States  (e-mail: se0201@princeton.edu)}
\address[Second]{Omenn-Darling Bioengineering Institute, Princeton University, United States}
\address[Third]{The Andlinger Center for Energy and the Environment, Princeton University, United States}
\address[Fourth]{High Meadows Environmental Institute, Princeton University, United States  (e-mail: javalos@princeton.edu)}

\begin{abstract}
We outline a modeling and optimization strategy for investigating dynamic metabolic engineering interventions. Our framework is particularly useful at the early stages of research and development, often constrained by limited knowledge and experimental data. Elucidating \textit{a priori} optimal trajectories of manipulatable intracellular fluxes can guide the design of suitable control schemes, e.g., cyber(ge)netic or in-cell approaches, and the selection of appropriate actuators, e.g., at the transcriptional or post-translational levels. Model-based dynamic optimization is proposed to predict optimal trajectories of target manipulatable intracellular fluxes. A challenge emerges as existing models are often oversimplified, lacking insights into metabolism, or excessively complex, making them difficult to build and implement. Here, we use surrogates derived from steady-state solutions of constraint-based metabolic models to link manipulatable intracellular fluxes to the process exchange rates of structurally simple \textit{hybrid} dynamic models. The latter can be conveniently used in optimal control problems of metabolism. As a proof of concept, we apply our method to a reduced metabolic network of \textit{Escherichia coli} considering two different scenarios of dynamic metabolic engineering.
\end{abstract}


\end{frontmatter}

\section{Introduction}
Microbial cell factories are engineered microorganisms optimized for biotechnological production. Metabolic engineering involves the rewiring of metabolic networks to enhance the production of native products or to introduce new (non-native) metabolic pathways into cells. As a result, microbial cell factories can make a wide range of biochemicals, biofuels, biomaterials, and biopharmaceuticals \citep{cho_designing_2022}. Bioprocess efficiency is commonly assessed by the product titer, volumetric productivity rate, and yield. In conventional metabolic engineering approaches, the cell's \textit{steady-state} metabolism is optimized to favor the product pathway. However, a well-known challenge is the fact that maximizing the product pathway diverts resources away from biomass synthesis, resulting in a trade-off where increased product yields lead to decreased biomass yields and reduced volumetric productivities \citep{venayak_engineering_2015}. 

Dynamic metabolic control \citep{venayak_engineering_2015} has emerged as a means to address intrinsic metabolic trade-offs. It focuses on dynamically adjusting manipulatable metabolic fluxes to enable temporal transitions between different metabolic states, rather than maintaining a static metabolic flux distribution. In contrast to exploiting only extracellular/process exchange rates as degrees of freedom, manipulating \textit{intracellular} fluxes can unlock more opportunities for optimizing the \textit{heart} of bioprocesses. The manipulation of target intracellular metabolic fluxes can be achieved, e.g., by tuning the expression of enzymes (transcriptional level) or by modifying enzyme activity (post-translational level) \citep{hoffman_optogenetics_2022}. In addition, control can be exerted externally (cybernetic control), utilizing external signals \citep{espinel_cybergenetic_2023}, or intracellularly (in-cell control), with control mechanisms encoded within the cell through genetic and molecular circuits \citep{dinh_development_2019}.

A key question arises: what should the optimal dynamic trajectories of the manipulatable fluxes be to maximize production efficiency? Knowing \textit{a priori} the optimal dynamic trajectories of the manipulatable intracellular fluxes can facilitate, e.g., the determination of whether a cybernetic or in-cell control approach is more suitable for implementation, especially valuable at early stages of research and development. Also, it could help to decide on the most appropriate actuation mechanism, e.g., either at the transcriptional or post-transcriptional level.  We propose employing model-based dynamic optimization to determine the optimal dynamic trajectories of target manipulatable intracellular fluxes that maximize production efficiency. To do so, we need a suitable dynamic model that links manipulatable intracellular metabolic fluxes to extracellular exchange rates such as substrate uptake, product excretion, and growth rates.

Unstructured and unsegregated macro-kinetic modeling may be the simplest way to model dynamic bioprocesses \citep{kyriakopoulos_kinetic_2018}. However, they do not capture information on intracellular metabolism; thus, not suitable for dynamic metabolic control applications exploiting \textit{intracellular} fluxes. Constraint-based modeling, such as flux balance analysis (FBA), enables the prediction of metabolic flux distributions in the steady state (cf. \cite{benner_stoichiometric_2014,klamt_mathematical_2018}). These (underdetermined) models optimize a cellular objective function. The optimization is constrained by the mass balances, constructed from the stoichiometric matrix of a metabolic network. Additional constraints related to, e.g., thermodynamics, regulation, and resource allocation, can also be applied.

Dynamic versions of constraint-based modeling, suitable for \textit{dynamic} optimization, are available (cf. e.g. \cite{waldherr_dynamic_2015,espinel_cybergenetic_2023}). Yet, integrating these models into process optimization turns the task into a bilevel optimization and requires assumptions on the cell's \textit{dynamic} objective function, e.g., short-term/long-term goals. Furthermore, the numerical solution of bilevel optimizations often involves game theory-based assumptions (cf. \cite{dempe_bilevel_2020}). One has to decide whether the two optimizations collaborate (optimistic approach) or conflict (pessimistic approach). The optimistic approach allows for the replacement of the inner optimization with its Karush–Kuhn–Tucker conditions. Nevertheless, this introduces non-convexity, even if the original model is convex. Consequently, constraint-based dynamic models present substantial challenges for process optimization. 

In this work, we employ neural networks to bridge the gap between the steady-state solutions of constraint-based models and macro-kinetic dynamic models. The neural network effectively creates a link between manipulatable intracellular fluxes and the process exchange rates. This methodology simplifies process optimization by circumventing bilevel optimization schemes while still enabling the use of manipulatable intracellular fluxes as optimization degrees of freedom. Section \ref{sec:metabolic_modeling} introduces our hybrid dynamic modeling strategy. In Section \ref{sec:dynamic_opti}, we present a model-based dynamic optimization scheme that exploits manipulatable intracellular fluxes to maximize production efficiency. As a proof of concept, Section \ref{sec:case_study} considers a reduced metabolic network of \textit{Escherichia coli} with selected intracellular fluxes as optimization degrees of freedom.

\section{Hybrid modeling strategy}
\label{sec:metabolic_modeling}
We proceed to introduce each of the elements of our proposed modeling framework.

\subsection{Constraint-based model in steady state}
\label{subsec:const_modeling}
A constraint-based model, often underdetermined, can be expressed as (\cite{klamt_mathematical_2018}):
\begin{maxi!} 
    {\bm{V}}{F_\mathrm{bio}(\bm{V}),\label{eq:FBA_cost}}{\label{eq:optimal}}{}
    \addConstraint{}{\dot{\bm{m}}=\bm{S}\bm{V} = \bm{0} \label{eq:SV_cons}}{}
    \addConstraint{}{\bm{V_\mathrm{min}} \leq \bm{V} \leq \bm{V_\mathrm{max}} \label{eq:V_cons}}{}
    \addConstraint{}{V_{i,\mathrm{min}} = 0, \, \forall i \in \mathbb{I} \label{eq:V_irr}}{}
    \addConstraint{}{\bm{V_\mathrm{man}} = \bm{v_\mathrm{man}}, \,\bm{V_\mathrm{man}} \subseteq \bm{V}  \label{eq:V_man_cons}}{}
    \addConstraint{}{\bm{0}\leq \bm{c}(\bm{V}). \label{eq:g_cons}}{}
\end{maxi!}

$\bm{S} \in \mathbb{R}^{n_m \times n_V}$ is the stoichiometric matrix associated with the internal metabolites $\bm{m} \in \mathbb{R}^{n_m}$ and metabolic fluxes $\bm{V} \in \mathbb{R}^{n_V}$. $\mathbb{I}$ represents the set of irreversible reactions, hence the lower bound is equal to zero. In case of manipulatable intracellular fluxes, denoted as $\bm{V_\mathrm{man}} \in \mathbb{R}^{n_\mathrm{man}}$, they can be constrained to their corresponding values $\bm{v_\mathrm{man}} \in \mathbb{R}^{n_\mathrm{man}}$. Additional (non-linear) constraints $\bm{c}: \mathbb{R}^{n_V} \rightarrow \mathbb{R}^{n_c}$ can be integrated to consider factors like resource allocation, thermodynamics, and regulation. The assumed cellular objective function is represented by $F_\mathrm{bio}:\mathbb{R}^{n_V} \rightarrow \mathbb{R}$, with $\bm{V}$ being the decision variable of the optimization. The model assumes steady-state conditions of the metabolism (cf. Eq. \eqref{eq:SV_cons}).

\subsection{Mapping intra- and extra-cellular metabolic fluxes}
\label{subsec:surrogate}
We categorize metabolic fluxes into intracellular fluxes, denoted as $\bm{V_\mathrm{int}} \in \mathbb{R}^{n_\mathrm{int}}$, and extracellular or exchange fluxes, represented by $\bm{V_\mathrm{ext}} \in \mathbb{R}^{n_\mathrm{ext}}$. Thus, $\bm{V} := [\bm{V_\mathrm{int}}^\tran,\bm{V_\mathrm{ext}}^\tran]^\tran$. For simplicity, we assume that all products and substrates are exchanged with the extracellular medium; however, there can also be intracellular products such as recombinant proteins. To systematically explore the impact on the metabolism of varying $\bm{V_\mathrm{man}}$, we employ a \textit{grid search} approach. We define sets $\mathbb{V}_{\mathrm{man},1}, \mathbb{V}_{\mathrm{man},2}, \ldots, \mathbb{V}_{\mathrm{man},n}$ representing possible values for each of the $n$ manipulatable intracellular fluxes. The Cartesian product gives us all possible combinations of these manipulatable flux values:
\begin{equation} 
\label{eq:cartesian_product}
\mathbb{G} = \prod_{i=1}^{n} \mathbb{V}_{\mathrm{man},i}.
\end{equation}
Each combination is applied as a constraint in Eq. \eqref{eq:V_man_cons}. For simplicity, we consider only one manipulatable intracellular flux in the case studies in Section \ref{sec:case_study}. Solving the constraint-based model in \eqref{eq:optimal} results in a set of \textit{labels} (extracellular/exchange fluxes) and \textit{features} (manipulatable intracellular fluxes). A machine-learning model can be used to learn the mapping between the intra- and extra-cellular flux domains. Note that, in principle, other modeling approaches can also be considered. Using neural networks, this can be expressed as:
\begin{equation}
\label{eq:ML_model_ext}
\bm{V}_\mathrm{ext} = \bm{f_\mathrm{NN}}(\bm{V}_\mathrm{man}, \bm{\Theta}),
\end{equation}
where $\bm{f_\mathrm{NN}}:\mathbb{R}^{n_\mathrm{man}} \times \mathbb{R}^{n_{\Theta}} \rightarrow \mathbb{R}^{n_\mathrm{ext}}$ represents the trained neural network and $\bm{\Theta} \in \mathbb{R}^{n_{\Theta}}$ represents the corresponding optimized parameters.

In cases of unfeasible combinations of manipulatable fluxes, i.e., where the FBA problem in \eqref{eq:optimal}  fails to converge, we propose to \textit{flag} these scenarios. For example, one could arbitrarily set the resulting fluxes equal to zero. Therefore, when Eq. \eqref{eq:ML_model_ext} is incorporated into model-based optimization problems (more details in Section \ref{sec:dynamic_opti}), the controller can be \textit{aware} of the \textit{flagged} unfeasible regions.

\textit{Remark}. We could optionally also consider selected (non-manipulatable) intracellular fluxes as labels of the surrogate of FBA, ensuring continued insight into the intracellular metabolic flux distribution.

\subsection{Hybrid dynamic model linked to intracellular fluxes}
\label{subsec:macro_augmented}
The neural-network surrogate described by Eq. \eqref{eq:ML_model_ext}, when integrated into macro-kinetic dynamic models, effectively creates a \textit{hybrid} machine-learning-supported \textit{dynamic} model. Therein, the process exchange rates are represented as functions of manipulatable intracellular fluxes. Let $\bm{z_\mathrm{ext}} \in \mathbb{R}^{n_z}$ be extracellular states of interest, including the biomass dry weight $z_\mathrm{ext,bio} \in \mathbb{R}$. Without loss of generality, let us assume a batch process. The dynamics of the system follows:
\begin{subequations} 
\begin{align} 
&\odv{\bm{z_\mathrm{ext}}(t)}{t} = z_\mathrm{ext,bio}(t) \cdot \bm{q}(\bm{V}_\mathrm{man}(t), \bm{\Theta}, \bm{z}(t), \bm{\theta}), \label{eq:z_dyn} \\
&q_{i} := V_{\mathrm{ext},{i}}(\bm{V}_{\mathrm{man}}(t),\bm{\Theta}) \cdot h(\bm{z}(t), \bm{\theta}) \label{eq:qi_def}, \\
&\forall i \in \{1,2,...,n_z\} \nonumber, \\
&\bm{z}(t_0) = \bm{z_0} \label{eq:z0}.
\end{align}
\end{subequations}

Here, $\bm{q}: \mathbb{R}^{n_{\mathrm{man}}} \times \mathbb{R}^{n_\Theta} \times \mathbb{R}^{n_z} \times \mathbb{R}^{n_{\theta}} \rightarrow \mathbb{R}^{n_z}$ describes the biomass-specific exchange reaction rates of the macro-kinetic model; $q_{i} \in \bm{q}$. $\bm{\theta} \in \mathbb{R}^{n_{\theta}}$ represents the parameters of the function  $h: \mathbb{R}^{n_z} \times \mathbb{R}^{n_{\theta}} \rightarrow \mathbb{R}$, while $t$ and $t_0$ denote the time and initial time, respectively. The function $h$ accounts for \textit{a priori} known rate-limiting factors, e.g., substrate uptake limitation or product inhibition, often neglected by steady-state constraint-based models. As such, $V_{\mathrm{ext},{i}}$ in Eq. \eqref{eq:ML_model_ext} represents \textit{maximum} theoretical flux values predicted by the constraint-based model. In the absence of (known) rate limitations, one could optimistically assume $h=1$ when, e.g., the carbon/electron source is available in the medium; otherwise, $h=0$. This simplification can facilitate the numerical integration of the dynamic system.  Henceforth, we will omit the time-dependency of variables when clear from the context.

\section{Dynamic metabolic control}
\label{sec:dynamic_opti}
We now show how the manipulatable intracellular fluxes can serve as degrees of freedom for metabolic control via model-based dynamic optimization.

\subsection{Dynamic optimization}
\label{subsec:open_loop}
The dynamic optimization problem for maximizing the production efficiency of a batch process reads:
\begin{maxi!}
    {\bm{V_\text{man}}(\cdot)|_{t_0}^{t_h}}{J_p(\bm{z},\bm{V_\text{man}},\bm{\theta},\bm{\Theta}),}{\label{eq:OLO_obj}}{}
    \addConstraint{\mathrm{Eqs.}\, \eqref{eq:ML_model_ext},\eqref{eq:z_dyn}-\eqref{eq:z0}}{}{\nonumber}
    \addConstraint{0 \leq \bm{g}(\bm{z},\bm{V_\text{man}},\bm{\theta},\bm{\Theta}),}{}{\label{eq:OLO_cons}}
\end{maxi!}
where $J_p: \mathbb{R}^{n_z} \times \mathbb{R}^{n_{\mathrm{man}}} \times \mathbb{R}^{n_{\theta}} \times \mathbb{R}^{n_{\Theta}} \rightarrow \mathbb{R}$ is the objective function (e.g., volumetric productivity, economic profit, a set-point, etc.). The function $\bm{g}: \mathbb{R}^{n_z} \times \mathbb{R}^{n_{\mathrm{man}}} \times \mathbb{R}^{n_{\theta}} \times \mathbb{R}^{n_{\Theta}} \rightarrow \mathbb{R}^{n_{g}}$ represents possible state and input constraints, addressing economic, technical, or safety considerations. In the case of trained surrogates with \textit{flagged} unfeasible flux regions resulting from specific (combinations of) manipulatable flux values, one can incorporate constraints into the optimizer via Eq. \eqref{eq:OLO_cons} to \textit{proactively} avoid these regions. 

The degree of freedom of the optimization problem is the \textit{function} of manipulatable fluxes $\bm{V_\text{man}}(\cdot)$ from $t_0$ to the final prediction time $t_h$, e.g., the final process time $t_f$ in a batch. The outcome of the process optimization problem can provide us with an idea of the \textit{theoretical} optimal trajectories of manipulatable fluxes (\textit{the path}) and aid in designing suitable actuation and control strategies (\textit{how to follow the path}).

\section{Proof of concept}
\label{sec:case_study}
\begin{figure}[htb]
    \begin{center}
        \includegraphics[scale=0.057]{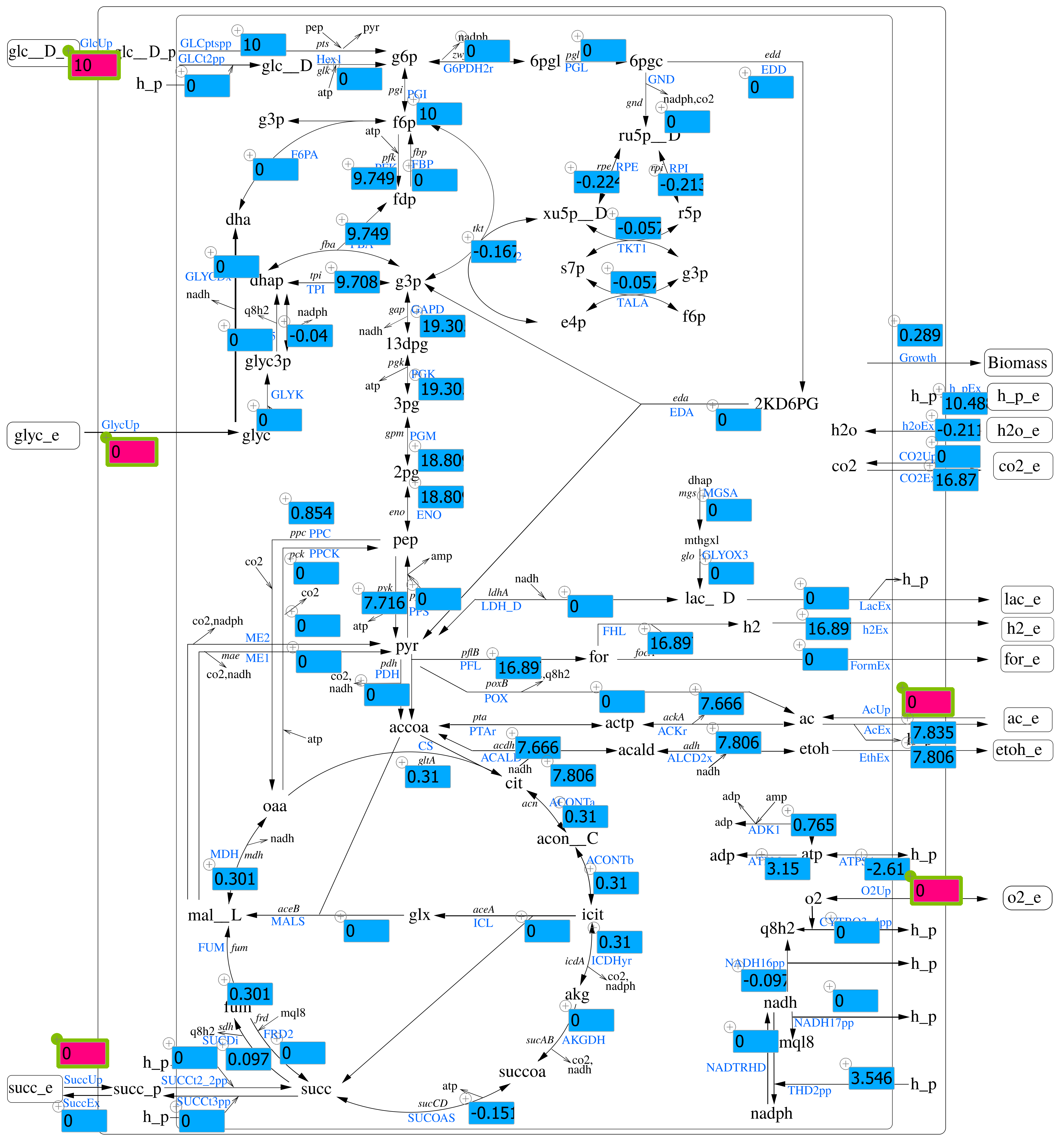}
        \caption{Example of an FBA simulation for the considered \textit{E. coli}'s metabolic network. FBA simulation and network map obtained with CNApy \citep{thiele_cnapy_2022}. The shown scenario belongs to a wild-type flux distribution of \textit{E. coli} under anaerobic conditions.} 
        \label{fig:network}
    \end{center}
\end{figure}

To outline the applicability of our proposed modeling and optimization framework, we use a compressed ECC2comp metabolic network of \textit{E. coli} \citep{hadicke_ecolicore2_2017}, retrieved from the available projects in the CNApy toolbox \citep{thiele_cnapy_2022}. The network (cf. Fig. \ref{fig:network}) captures \textit{E. coli}'s central metabolism, comprising 122 reactions and 94 metabolites. We consider three test scenarios which will be detailed in the following sections.

\textit{Remark on computational methods}. When applicable, we consider piece-wise constant manipulatable intracellular fluxes of size $0.01\, \mathrm{h}$ to approximate the continuous \textit{function} $\bm{V_\text{man}}(\cdot)
    $, otherwise the process optimization problems are infinite-dimensional. COBRApy \citep{ebrahim_cobrapy_2013} is used to systematically solve the constraint-based model (FBA) and populate the dataset of labels and features for training the surrogates. The training of the neural networks and the neural-network-supported dynamic optimizations are conducted in HILO-MPC \citep{pohlodek_flexible_2024}. Initial conditions: $z_\mathrm{ext,bio}(t_0)=0.01 \, \mathrm{g/L}$ and $z_\mathrm{ext,glc}(t_0)=10 \, \mathrm{mmol/L}$; the other states were set to zero.

\textit{Remark on FBA}. In all scenarios, the assumed cell's objective function in \eqref{eq:FBA_cost} is set to the maximization of the biomass reaction flux. We consider \textit{anaerobic conditions} (oxygen uptake flux set to zero) and an upper bound for the glucose uptake of $10 \, \mathrm{mmol/g/h}$. Gene knock-outs are simulated by setting the corresponding fluxes to zero.

\textit{Remark on the surrogate training}. In all scenarios, 1,000 equidistant flux values were used to populate the corresponding sets of manipulatable fluxes (cf. Eq. \eqref{eq:cartesian_product}). For training the neural networks (Eq. \eqref{eq:ML_model_ext}), 20 \% of the data was used for testing; 80 \% of the remaining data was used for training and 20 \% for validation (early stopping).

\textit{Remark on units}. We express the extracellular states in $\mathrm{mmol/L}$, except for biomass which is expressed in $\mathrm{g_b/L}$, where $\mathrm{g_b}$ is the cell dry weight. The metabolic fluxes are biomass-specific, expressed in $\mathrm{mmol/g_b/h}$, except for the growth rate which is in $\mathrm{1/h}$.

\subsection{Scenario 1}
\label{subsec:scenario_1}
\begin{figure}[htb!]
    \begin{center}
        \vspace{-0.2cm}
        \includegraphics[trim={0 0 0 1.2cm},clip,scale=0.7]{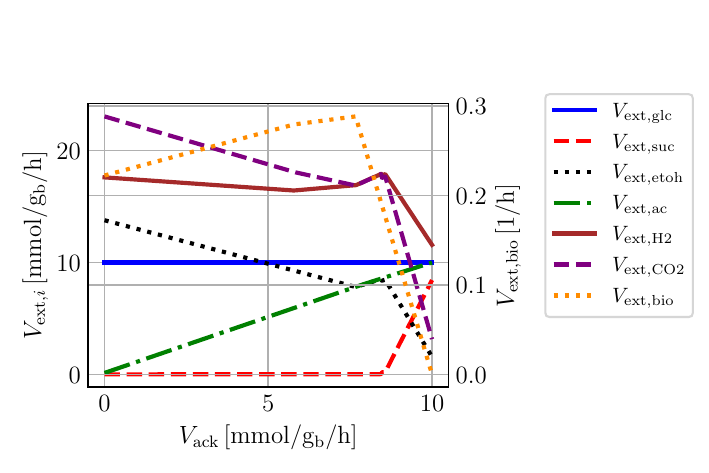}
        \vspace{-0.3cm}
        \caption{Explored flux space used to build the surrogate model in scenario 1.} 
        \label{fig:flux_space_scenario_1}
    \end{center}
\end{figure}

In this scenario, the focus is on the hybrid modeling strategy.  We consider the dynamic manipulation of the flux responsible for acetate synthesis (acetate kinase, \textit{ack}) due to its significant effect on the cell's energy state and metabolic flux distribution, thus $\bm{V_\mathrm{man}}:=V_\mathrm{ack}$. The extracellular metabolic fluxes resulting from the systematic exploration of the manipulatable flux (cf. Section \ref{subsec:surrogate}) are shown in Fig. \ref{fig:flux_space_scenario_1}. It covers the manipulatable flux range starting from zero until the upper bound for which the FBA problem was still feasible, $[0,10] \, \mathrm{mmol/g_b/h}$. As expected, the acetate kinase flux ($V_\mathrm{ack}$) directly influences the acetate exchange flux ($V_{\mathrm{ext,ac}}$). The reaction catalyzed by acetate kinase produces ATP which can be used to sustain growth. Therefore, an increasing trend of biomass growth rate ($V_{\mathrm{ext,bio}}$) is observed with increasing $V_\mathrm{ack}$ until an optimal value for growth is reached at $V_\mathrm{ack} \approx 7.66 \, \mathrm{mmol/g_b/h}$, matching the predicted wild-type phenotype.  However, even though the acetate exchange flux can be increased further by incrementing $V_\mathrm{ack}$, the resulting flux distribution (the intricate carbon, energy, and redox balancing) turns unfavorable for growth.

Furthermore, the ethanol exchange flux ($V_{\mathrm{ext,etoh}}$) is highest at $V_\mathrm{ack} = 0$ due to the higher carbon flux available, and overall decreases with increasing $V_\mathrm{ack}$ as more carbon flux goes to acetate. The relationship between fluxes for ethanol, acetate, and growth highlights an area of opportunity in metabolic engineering. For example, the ethanol volumetric productivity can be maximized by dynamically modulating $V_\mathrm{ack}$, which has been discussed by \cite{jabarivelisdeh_improving_2016}. Thereby, one can have a \textit{growth phase} with $V_\mathrm{ack}$ active and low ethanol production, followed by a \textit{production phase} with $V_\mathrm{ack}$ inactive and high ethanol production. This will be further discussed in scenario 2. Note that the succinate exchange flux ($V_{\mathrm{ext,suc}}$) is predicted to be active at high values of $V_\mathrm{ack}$ to enable redox balancing, which also correlates to reduced ethanol fluxes. The glucose exchange flux ($V_{\mathrm{ext,glc}}$) takes the maximum allowed value in all cases. Fig. \ref{fig:flux_space_scenario_1} also shows the exchange flux values of gases CO\textsubscript{2} ($V_{\mathrm{ext,CO2}}$) and H\textsubscript{2} ($V_{\mathrm{ext,H2}}$) with changes in $V_\mathrm{ack}$.

\begin{figure}[htb!]
    \begin{center}    
        \subfigure[$V_{\mathrm{ext,bio}}$]{\includegraphics[scale=0.48]{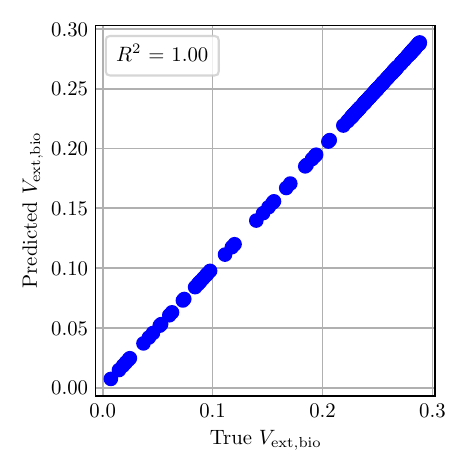}}
        \subfigure[$V_{\mathrm{ext,etoh}}$]{\includegraphics[scale=0.48]{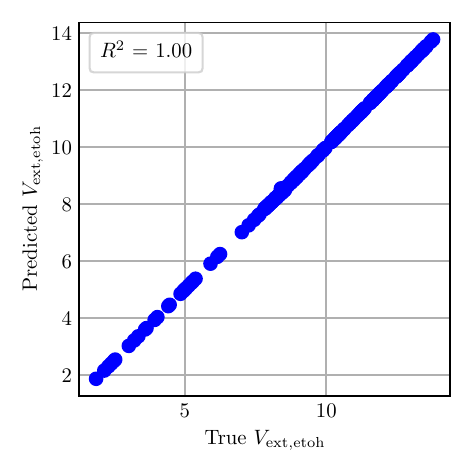}}
        \vspace{-0.3cm}
        \caption{\textit{Selected} parity plots of the trained neural-network surrogate of FBA (scenario 1) for the (a) biomass and (b) ethanol extracellular fluxes.}
        \label{fig:parity}
    \end{center}
\end{figure}

We trained a neural network to obtain a surrogate model as in Eq. \eqref{eq:ML_model_ext}, considering all the exchange fluxes in Fig. \ref{fig:flux_space_scenario_1} as labels, except for the constant $V_{\mathrm{ext,glc}}$. The neural network consisted of one hidden layer, five neurons, and rectified linear unit (ReLU) activation functions. The coefficient of determination for the parity plots of all exchange fluxes, computed with the testing dataset, was $R^2=1.00$ (cf. Fig. \ref{fig:parity}), indicating that the surrogate was able to successfully learn the solution of the FBA model. 

\begin{figure}[htb!]
    \begin{center}
        \subfigure[Manipulated $V_\mathrm{ack}$]{\includegraphics[scale=0.48]{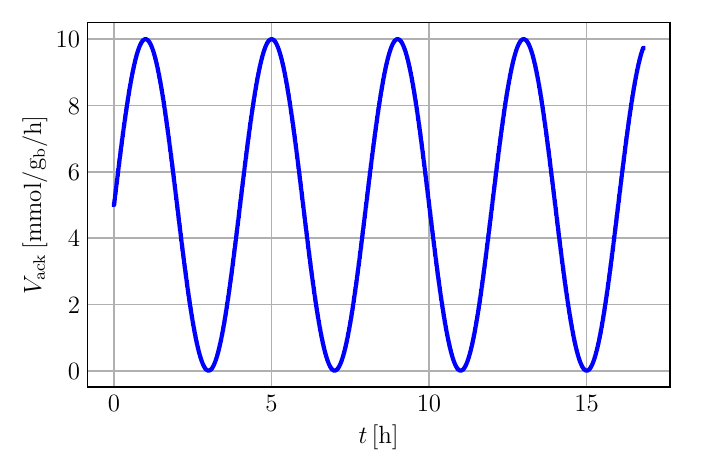}}
        \vspace{-0.5cm}
        \subfigure[Fermentation dynamics]{\includegraphics[trim={0 0 0 1.3cm},clip,scale=0.71]{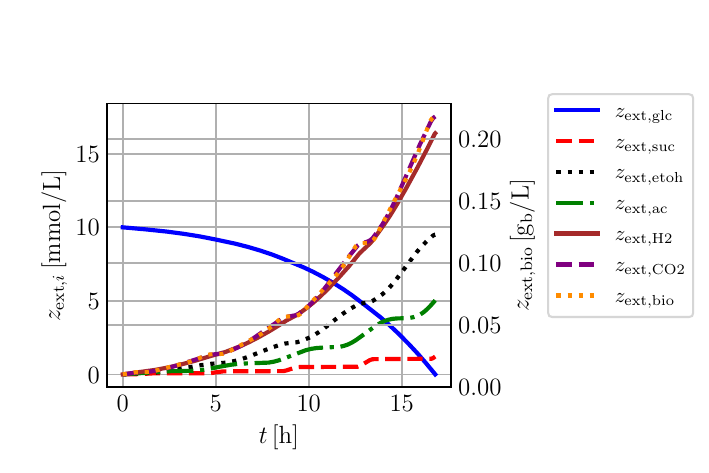}}
        \caption{Effect of (a) a given trajectory of $V_\mathrm{ack}$ and (b) the predicted dynamic profiles of relevant extracellular states in scenario 1.}
        \label{fig:base_case_dynamics}
    \end{center}
\end{figure}

The hybrid model was built as in Eqs. \eqref{eq:z_dyn}-\eqref{eq:z0}. The function $h$ was modeled as:
\begin{equation}
    h = \left(\frac{z_{\mathrm{ext,glc}}}{z_{\mathrm{ext,glc}}+k_s}\right) \cdot \left(\frac{1}{1+z_{\mathrm{ext,etoh}}/k_i}\right),
\end{equation}
where $k_s = 2.964\times10^{-4} \, \mathrm{mmol/L}$ \citep{senn_growth_1994} and $k_i = 25 \, \mathrm{mmol/L}$ \citep{gadkar_estimating_2005} are glucose saturation and ethanol inhibition constants, respectively. Considering an arbitrary $V_\mathrm{ack}$ sinusoidal signal input, we show in Fig. \ref{fig:base_case_dynamics} the predicted fermentation dynamics. The signal oscillates in the range of $[0,10] \, \mathrm{mmol/g_b/h}$. As expected, the dynamic behavior of the states coincided with the increasing and decreasing trends of the metabolic exchange fluxes as a function of $V_\mathrm{ack}$ presented in Fig \ref{fig:flux_space_scenario_1}.

\subsection{Scenario 2}
\label{subsec:scenario_2}
We now assume that one would like to maximize ethanol volumetric productivity. To do so, we consider knock-outs of lactate dehydrogenase (\textit{ldhA}) and fumarate reductase (\textit{frdA}), effectively disabling lactate and succinate synthesis. Furthermore, instead of knocking out acetate kinase to increase the flux to ethanol, we aim to determine an optimal flux manipulation strategy of $V_\mathrm{ack}$, similar to \cite{jabarivelisdeh_improving_2016}. The idea is to \textit{balance} growth (linked to acetate synthesis) and ethanol production optimally to maximize ethanol volumetric productivity, as discussed in scenario 1. To model this system, we followed similar steps as with scenario 1. Here, the neural-network surrogate consisted of one hidden layer, four neurons, and sigmoid activation functions; the labels are the acetate, ethanol, and growth extracellular fluxes. We solved an optimization problem constrained by the hybrid model to find the optimal trajectory of $V_\mathrm{ack}$ that maximizes the final ethanol concentration in the considered time frame:
\begin{maxi!}
    {\bm{V_\text{ack}}(\cdot)|_{t_0}^{t_f}}{z_{\mathrm{ext,etoh}}(t_f),}{}{}
    \addConstraint{\mathrm{Eqs.}\, \eqref{eq:ML_model_ext},\eqref{eq:z_dyn}-\eqref{eq:z0}.}{}{\nonumber}
\end{maxi!}

\begin{figure}[htb!]
    \begin{center}
        \subfigure[Manipulated $V_\mathrm{ack}$]{\includegraphics[scale=0.48]{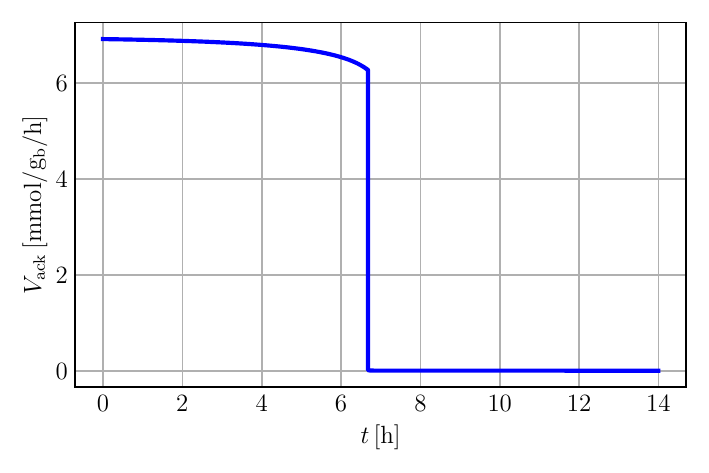}}
        \vspace{-0.5cm}
        \subfigure[Fermentation dynamics]{\includegraphics[trim={0 0 0 1.3cm},clip,scale=0.71]{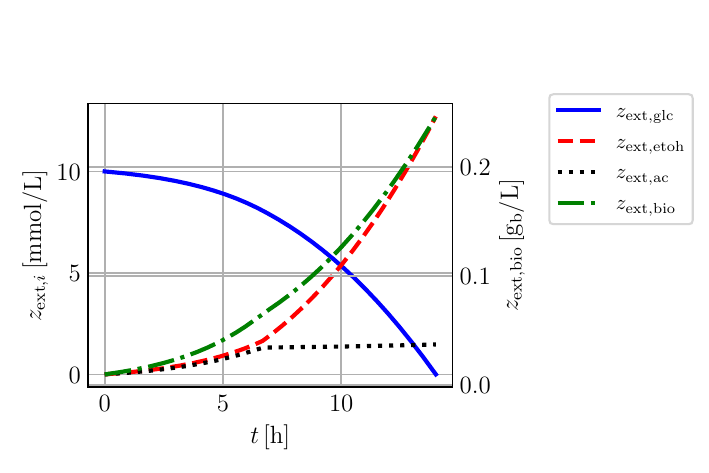}}
        \caption{Effect of (a) an optimized trajectory of $V_\mathrm{ack}$ and (b) the predicted dynamic profiles of relevant extracellular states in scenario 2.}
        \label{fig:case_2_dynamics}
    \end{center}
\end{figure}

The predicted behavior of the optimized system is shown in Fig. \ref{fig:case_2_dynamics}. As expected, a two-stage fermentation is predicted that maximizes the volumetric productivity of ethanol. Initially, there is a growth phase at a $V_\mathrm{ack}$ such that it ensures maximum growth, with little ethanol production, followed by the inactivation of $V_\mathrm{ack}$, which halts growth but enhances ethanol synthesis.

\subsection{Scenario 3}
\label{subsec:scenario_3}
\label{subsec:ATP_turnover}
As a last example, we consider the dynamic manipulation of the ATP turnover to enhance lactate yield \citep{espinelrios_maximizing_2022,espinel_cybergenetic_2023,espinel-rios_experimentally_2024}. Enforcing an ATP turnover under these conditions (i.e., product synthesis linked to ATP generation) is expected to increase the lactate flux to compensate for the loss of ATP, although at the expense of growth \citep{espinelrios_maximizing_2022,espinel_cybergenetic_2023,espinel-rios_experimentally_2024}. Therefore, the challenge is to optimally \textit{balance} the product-biomass trade-off during the fermentation. To simulate ATP turnover, an ATP-\textit{wasting} reaction is introduced into the network. The associated flux is to be dynamically manipulated, $\bm{V_\mathrm{man}}:=V_\mathrm{atpAGD}$. To have lactate as the main fermentation product, we assume that the acetate- and ethanol-producing reactions are blocked by simulating deletions of genes \textit{pta}/\textit{ack} and \textit{adhE}, respectively. To model this system, we followed similar steps as with scenario 1. Here, the surrogate consists of one hidden layer, one neuron, and a sigmoid activation function; the labels are only the lactate and growth extracellular fluxes. We solved an optimization problem constrained by the hybrid model to find the optimal trajectory of $V_\mathrm{atpAGD}$ that maximizes the final lactate concentration in the considered time frame:
\begin{maxi!}
    {\bm{V_\text{atpAGD}}(\cdot)|_{t_0}^{t_f}}{z_{\mathrm{ext,lac}}(t_f),}{}{}
    \addConstraint{\mathrm{Eqs.}\, \eqref{eq:ML_model_ext},\eqref{eq:z_dyn}-\eqref{eq:z0}.}{}{\nonumber}
\end{maxi!}

\begin{figure}[htb!]
    \begin{center}
        \subfigure[Manipulated $V_\mathrm{atpAGD}$]{\includegraphics[scale=0.48]{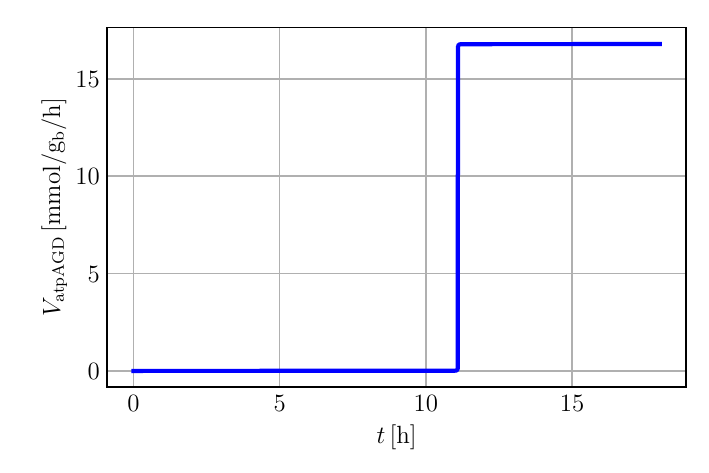}}
        \vspace{-0.5cm}
        \subfigure[Fermentation dynamics]{\includegraphics[trim={0 0 0 1.3cm},clip,scale=0.71]{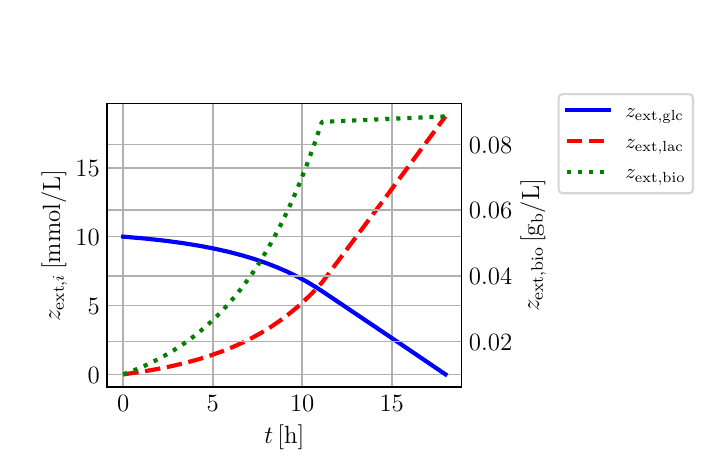}}
        \caption{Effect of (a) an optimized trajectory of $V_\mathrm{atpAGD}$ and (b) the predicted dynamic profiles of relevant extracellular states in scenario 3.} 
        \label{fig:case_3_dynamics}
    \end{center}
\end{figure}

The predicted behavior of the optimized system is shown in Fig. \ref{fig:case_3_dynamics}. As expected, the optimizer predicts a first phase with $V_\mathrm{atpAGD}$ inactive to ensure maximum growth (\textit{growth phase}), followed by full activation of $V_\mathrm{atpAGD}$ (\textit{production phase}) with enhanced lactate synthesis.

\section{Conclusion}
This study bridges manipulatable intracellular metabolic fluxes with process exchange rates. Surrogate models are trained to learn the solutions of a constraint-based metabolic model (FBA). A simple \textit{hybrid} \textit{dynamic} model is then formulated, whereby the exchange process rates depend on the manipulatable intracellular fluxes via the surrogates. This approach not only enables predictions of the dynamic extracellular states given changes in target intracellular fluxes, but also facilitates dynamic optimization using intracellular fluxes as optimization degrees of freedom. As demonstrated with the selected case studies, the proposed approach facilitates efficient \textit{in silico} testing of diverse dynamic metabolic engineering strategies at the flux level. As such, it can aid in the design of suitable control and actuation mechanisms of metabolism. This is especially beneficial when constrained by limited knowledge of the system, frequently in early research and development.

\bibliography{ifacconf}

\begin{thebibliography}{19}
\providecommand{\natexlab}[1]{#1}
\providecommand{\url}[1]{\texttt{#1}}
\providecommand{\urlprefix}{URL }
\expandafter\ifx\csname urlstyle\endcsname\relax
  \providecommand{\doi}[1]{doi:\discretionary{}{}{}#1}\else
  \providecommand{\doi}{doi:\discretionary{}{}{}\begingroup \urlstyle{rm}\Url}\fi

\bibitem[{Cho et~al.(2022)Cho, Kim, Eun, Moon, and Lee}]{cho_designing_2022}
Cho, J.S., Kim, G.B., Eun, H., Moon, C.W., and Lee, S.Y. (2022).
\newblock Designing microbial cell factories for the production of chemicals.
\newblock \emph{JACS Au}, 2(8), 1781--1799.

\bibitem[{Dempe(2020)}]{dempe_bilevel_2020}
Dempe, S. (2020).
\newblock Bilevel optimization: theory, algorithms, applications and a bibliography.
\newblock In S.~Dempe and A.~Zemkoho (eds.), \emph{Bilevel {Optimization}}, volume 161, 581--672. Springer International Publishing, Cham.

\bibitem[{Dinh and Prather(2019)}]{dinh_development_2019}
Dinh, C.V. and Prather, K.L.J. (2019).
\newblock Development of an autonomous and bifunctional quorum-sensing circuit for metabolic flux control in engineered \textit{{Escherichia} coli}.
\newblock \emph{Proc Natl Acad Sci USA}, 116(51), 25562--25568.

\bibitem[{Ebrahim et~al.(2013)Ebrahim, Lerman, Palsson, and Hyduke}]{ebrahim_cobrapy_2013}
Ebrahim, A., Lerman, J.A., Palsson, B.O., and Hyduke, D.R. (2013).
\newblock {COBRApy}: {COnstraints}-{Based} {Reconstruction} and {Analysis} for {Python}.
\newblock \emph{BMC Syst Biol}, 7(1), 74.

\bibitem[{Espinel-Ríos et~al.(2024)Espinel-Ríos, Behrendt, Bauer, Morabito, Pohlodek, Schütze, Findeisen, Bettenbrock, and Klamt}]{espinel-rios_experimentally_2024}
Espinel-Ríos, S., Behrendt, G., Bauer, J., Morabito, B., Pohlodek, J., Schütze, A., Findeisen, R., Bettenbrock, K., and Klamt, S. (2024).
\newblock Experimentally implemented dynamic optogenetic optimization of {ATPase} expression using knowledge-based and {Gaussian}-process-supported models.
\newblock In arXiv:2401.08556.

\bibitem[{Espinel‐Ríos et~al.(2022)Espinel‐Ríos, Bettenbrock, Klamt, and Findeisen}]{espinelrios_maximizing_2022}
Espinel‐Ríos, S., Bettenbrock, K., Klamt, S., and Findeisen, R. (2022).
\newblock Maximizing batch fermentation efficiency by constrained model‐based optimization and predictive control of adenosine triphosphate turnover.
\newblock \emph{AIChE J}, 68(4), e17555.

\bibitem[{Espinel‐Ríos et~al.(2024)Espinel‐Ríos, Morabito, Pohlodek, Bettenbrock, Klamt, and Findeisen}]{espinel_cybergenetic_2023}
Espinel‐Ríos, S., Morabito, B., Pohlodek, J., Bettenbrock, K., Klamt, S., and Findeisen, R. (2024).
\newblock Toward a modeling, optimization, and predictive control framework for fed‐batch metabolic cybergenetics.
\newblock \emph{Biotechnol Bioeng}, 121(1), 366--379.

\bibitem[{Gadkar et~al.(2005)Gadkar, Doyle~Iii, Edwards, and Mahadevan}]{gadkar_estimating_2005}
Gadkar, K.G., Doyle~Iii, F.J., Edwards, J.S., and Mahadevan, R. (2005).
\newblock Estimating optimal profiles of genetic alterations using constraint-based models.
\newblock \emph{Biotechnol Bioeng}, 89(2), 243--251.

\bibitem[{Hoffman et~al.(2022)Hoffman, Tang, and Avalos}]{hoffman_optogenetics_2022}
Hoffman, S.M., Tang, A.Y., and Avalos, J.L. (2022).
\newblock Optogenetics illuminates applications in microbial engineering.
\newblock \emph{Annu Rev Chem Biomol}, 13(1), 373--403.

\bibitem[{Hädicke and Klamt(2017)}]{hadicke_ecolicore2_2017}
Hädicke, O. and Klamt, S. (2017).
\newblock {EColiCore2}: a reference network model of the central metabolism of \textit{Escherichia coli} and relationships to its genome-scale parent model.
\newblock \emph{Sci Rep}, 7(1), 39647.

\bibitem[{Jabarivelisdeh and Waldherr(2016)}]{jabarivelisdeh_improving_2016}
Jabarivelisdeh, B. and Waldherr, S. (2016).
\newblock Improving bioprocess productivity using constraint-based models in a dynamic optimization scheme.
\newblock \emph{IFAC-PapersOnLine}, 49(26), 245--251.

\bibitem[{Klamt et~al.(2014)Klamt, Hädicke, and Von~Kamp}]{benner_stoichiometric_2014}
Klamt, S., Hädicke, O., and Von~Kamp, A. (2014).
\newblock Stoichiometric and constraint-based analysis of biochemical reaction networks.
\newblock In P.~Benner, R.~Findeisen, D.~Flockerzi, U.~Reichl, and K.~Sundmacher (eds.), \emph{Large-{Scale} {Networks} in {Engineering} and {Life} {Sciences}}, 263--316. Springer International Publishing, Cham.

\bibitem[{Klamt et~al.(2018)Klamt, Müller, Regensburger, and Zanghellini}]{klamt_mathematical_2018}
Klamt, S., Müller, S., Regensburger, G., and Zanghellini, J. (2018).
\newblock A mathematical framework for yield (vs. rate) optimization in constraint-based modeling and applications in metabolic engineering.
\newblock \emph{Metab Eng}, 47, 153--169.

\bibitem[{Kyriakopoulos et~al.(2018)Kyriakopoulos, Ang, Lakshmanan, Huang, Yoon, Gunawan, and Lee}]{kyriakopoulos_kinetic_2018}
Kyriakopoulos, S., Ang, K.S., Lakshmanan, M., Huang, Z., Yoon, S., Gunawan, R., and Lee, D. (2018).
\newblock Kinetic modeling of mammalian cell culture bioprocessing: the quest to advance biomanufacturing.
\newblock \emph{Biotechnol J}, 13(3), 1700229.

\bibitem[{Pohlodek et~al.(2024)Pohlodek, Morabito, Schlauch, Zometa, and Findeisen}]{pohlodek_flexible_2024}
Pohlodek, J., Morabito, B., Schlauch, C., Zometa, P., and Findeisen, R. (2024).
\newblock Flexible development and evaluation of machine‐learning‐supported optimal control and estimation methods via {HILO}‐{MPC}.
\newblock \emph{Int J Robust Nonlin}, rnc.7275.

\bibitem[{Senn et~al.(1994)Senn, Lendenmann, Snozzi, Hamer, and Egli}]{senn_growth_1994}
Senn, H., Lendenmann, U., Snozzi, M., Hamer, G., and Egli, T. (1994).
\newblock The growth of \textit{{Escherichia} coli} in glucose-limited chemostat cultures: a re-examination of the kinetics.
\newblock \emph{Biochim Biophys Acta Gen Subj}, 1201(3), 424--436.

\bibitem[{Thiele et~al.(2022)Thiele, Von~Kamp, Bekiaris, Schneider, and Klamt}]{thiele_cnapy_2022}
Thiele, S., Von~Kamp, A., Bekiaris, P.S., Schneider, P., and Klamt, S. (2022).
\newblock {CNApy}: a {CellNetAnalyzer} {GUI} in {Python} for analyzing and designing metabolic networks.
\newblock \emph{Bioinformatics}, 38(5), 1467--1469.

\bibitem[{Venayak et~al.(2015)Venayak, Anesiadis, Cluett, and Mahadevan}]{venayak_engineering_2015}
Venayak, N., Anesiadis, N., Cluett, W.R., and Mahadevan, R. (2015).
\newblock Engineering metabolism through dynamic control.
\newblock \emph{Curr Opin Biotechnol}, 34, 142--152.

\bibitem[{Waldherr et~al.(2015)Waldherr, Oyarzún, and Bockmayr}]{waldherr_dynamic_2015}
Waldherr, S., Oyarzún, D.A., and Bockmayr, A. (2015).
\newblock Dynamic optimization of metabolic networks coupled with gene expression.
\newblock \emph{J Theor Biol}, 365, 469--485.

\end{thebibliography}

\end{document}